\documentclass[10pt,twocolumn,letterpaper]{article}

\usepackage{iccv}
\usepackage{times}
\usepackage{epsfig}
\usepackage{graphicx}
\usepackage{amsmath}
\usepackage{amssymb}
\usepackage{xcolor}
\usepackage{subfig}
\usepackage{physics}
\usepackage{booktabs}
\usepackage{graphicx}
\usepackage{textcomp}
\usepackage{algorithmic}
\usepackage{filecontents}
\usepackage{lipsum}
\usepackage{mathtools}
\usepackage{cuted}
\usepackage{tabularx}
\usepackage[accsupp]{axessibility}
\usepackage[pagebackref=true,breaklinks=true,letterpaper=true,colorlinks,bookmarks=false]{hyperref}

\usepackage[format=plain,labelfont=it,labelfont=bf,labelsep=period,textfont=it]{caption}
\iccvfinalcopy
\def\iccvPaperID{9595}

\definecolor{blue_color}{rgb}{0.0, 0.0, 1.0}
\definecolor{red_color}{rgb}{1.0, 0.0, 0.0}

\newcommand{\sol}{MosaiQ}
\newcommand{\sota}{QGPatch}

\begin{document}

\title{\vspace{-8mm}\sol{}: Quantum Generative Adversarial Networks \\for Image Generation on NISQ Computers\vspace{0mm}}
\author{Daniel Silver\\
Northeastern University\\
{\tt\small silver.da@northeastern.edu}
\and
Tirthak Patel\\
Rice University\\
{\tt\small tirthak.patel@rice.edu}
\and
William Cutler\\
Northeastern University\\
{\tt\small cutler.wi@northeastern.edu}
\and
Aditya Ranjan\\
Northeastern University\\
{\tt\small ranjan.ad@northeastern.edu}
\and
Harshitta Gandhi\\
Northeastern University\\
{\tt\small gandhi.ha@northeastern.edu}
\and
Devesh Tiwari\\
Northeastern University\\
{\tt\small d.tiwari@northeastern.edu}
}
% \author{Daniel Silver
% \and Tirthak Patel\\
% \and William Cutler\\
% \and Aditya Ranjan\\
% \and Harshitta Gandhi\\
% \and Devesh Tiwari\\
% }

\maketitle
\ificcvfinal\thispagestyle{empty}\fi

\begin{abstract}
    Quantum machine learning and vision have come to the fore recently, with hardware advances enabling rapid advancement in the capabilities of quantum machines. Recently, quantum image generation has been explored with many potential advantages over non-quantum techniques; however, previous techniques have suffered from poor quality and robustness.  To address these problems, we introduce \sol{} a high-quality quantum image generation GAN framework that can be executed on today's Near-term Intermediate Scale Quantum (NISQ) computers.\footnote{Accepted to appear in  the proceedings of International Conference on Computer Vision (ICCV), 2023. This is authors' pre-print copy.}
\end{abstract}

\section{Introduction}

Generative Adversarial Networks, or GANs, are a type of neural network architecture used in machine learning and computer vision for generative modeling~\cite{firstGAN, https://doi.org/10.48550/arxiv.1703.10593, https://doi.org/10.48550/arxiv.1701.07875, https://doi.org/10.48550/arxiv.1704.00028}. A classical GAN consists of two neural networks, a generator, and a discriminator, that are trained simultaneously in a competitive process. The generator generates fake data samples, while the discriminator tries to distinguish them from real ones found in the training set, hence serving as an ``adversarial entity''. Classical GANs have received significant attention for generating high-quality images, among other purposes including text generation, data augmentation, and anomaly detection \cite{XIA2022497,10.1007/978-3-030-20893-6_39,deecke2018anomaly}.

Naturally, this has spawned interest in the quantum information science community to develop corresponding quantum GANs that run on quantum computers. While recent efforts toward developing quantum GANs have been instrumental and early results have been encouraging, we discovered that existing approaches have severe scalability bottlenecks and have significant room for improvement. This is especially true in the context of generating high-quality image generation on real-system quantum computers.

\subsection*{Opportunity Gap for Quantum GANs.}

A recent work by Huang et al.~\cite{huang2021experimental}, referred to as \sota{} in this paper, is the state-of-the-art demonstration of QuantumGANs on real quantum computers. As our paper also demonstrates, \sota{} can learn different shapes and produce recognizable images in some cases, but can often yield low-quality images. It suffers from the scalability challenge because  it  breaks the image into ``patches'' and performs a pixel-by-pixel learning. Second, it is not effective at generating a variety of images within the same class -- this problem is known as ``mode collapse''~\cite{problemsGAN}. It is non-trivial to achieve high-quality image generation while also maintaining variety. Motivated by these limitations, \sol{}'s design pushes the state of the art by achieving higher scalability, image quality, and variety.

\subsection*{Contributions of \sol{}}

\noindent \textbf{I.} \sol{} introduces the design and implementation of a novel \textit{quantum generative adversarial network for image generation on quantum computers}. \sol{}'s approach is a hybrid classical-quantum generation network where a network of low-circuit-depth variational quantum circuits are leveraged to learn and train the model. Upon acceptance, \sol{} will be available as an open-source contribution.

\vspace{2mm}

\noindent \textbf{II.} \sol{}'s design demonstrates how the extraction of principal components of images enables us to learn and generate higher-quality images, compared to the state-of-the-art approach which is limited in its scalability due to pixel-by-pixel learning~\cite{huang2021experimental}. However, exploiting information in principal components to its full potential is non-trivial, and \sol{} proposes to mitigate those challenges using feature redistribution. Furthermore, \sol{} introduces a novel adaptive input noise generation technique to improve both the quality and variety of generated images -- mitigating the common risk of mode collapse in generative networks.

\vspace{5mm}

\noindent \textbf{III.} Our evaluation demonstrates that \sol{} significantly outperforms the state-of-the-art methods~\cite{huang2021experimental} on both simulation and real quantum computers with a hardware error. \sol{} is evaluated on the MNIST \cite{mnist} and Fashion MNIST datasets~\cite{xiao2017/online} -- widely-used for QML evaluation on Near-term Intermediate Scale Quantum (NISQ) machines~\cite{nakaji2021quantum, huang2021experimental}. \sol{} outperforms the state-of-the-art methods~\cite{huang2021experimental} both visually and quantitatively -- \textit{for example, over 100 points improvement in image quality generated on IBM Jakarta quantum computer using the FID (Fréchet Inception Distance) score~\cite{heusel2017gans}}, which is popularly used for comparing image quality. 

\section{\sol{}: Challenges and Solution}

\begin{figure}
    \centering
    \includegraphics[scale=0.37]{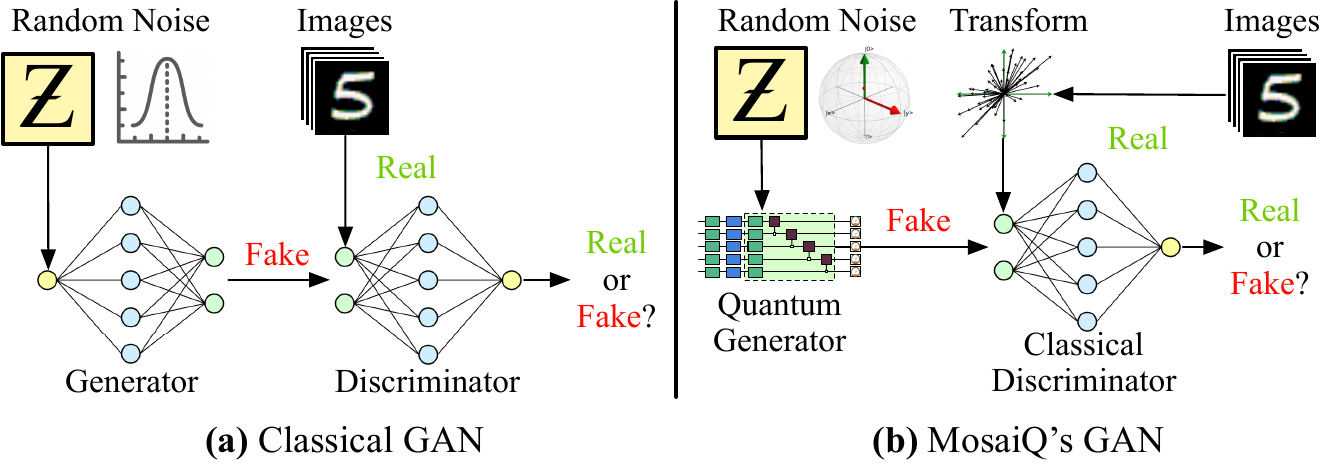}
    \vspace{1mm}
    \hrule
    \vspace{1mm}
    \caption{Classical generative adversarial network (GAN) and \sol{}'s hybrid quantum-classical GAN architecture.}
    \label{fig:classical_and_quantum_gan}
\end{figure}

In this section, we present the design and implementation of \sol{}, a quantum image generative network. To provide better context for understanding, first, we briefly describe the classical generative adversarial networks (GAN) (depicted in Fig.~\ref{fig:classical_and_quantum_gan}(a)), describe the limitations of the state-of-the-art quantum GANs, and then, provide the details of \sol{}'s quantum GAN design (depicted in Fig.~\ref{fig:classical_and_quantum_gan}(b)).

\subsection{Generative Adversarial Networks (GANs)}

Generative Adversarial Networks, or GANs, are a type of neural network architecture used in machine learning for generative modeling \cite{firstGAN}. The basic architecture and workflow of a classical (non quantum)  GAN are shown in Fig.~\ref{fig:classical_and_quantum_gan} (a). A classical GAN consists of two neural networks, a generator, and a discriminator, that are trained simultaneously in a competitive process. The generator generates fake data samples, while the discriminator tries to distinguish them from real ones found in the training set, hence serving as the ``adversary''. Through training, the generator learns to create increasingly realistic samples that mimic the distribution of the real data in order to better fool an increasingly effective discriminator, eventually reaching an attempt to attain equilibrium with the value function expressed as $\min _G \max _D \mathbb{E}_{x \sim q_{\mathrm{data}}(\boldsymbol{x})}[\log D(\boldsymbol{x})] + \mathbb{E}_{\boldsymbol{z} \sim p(\boldsymbol{z})}[\log (1-D(G(\boldsymbol{z})))]$. This value function has two main identifiable components that correspond to the respective optimization objectives of the generator and the discriminator. The use of $\log$ provides more numerical stability because it converts the multiplication of multiple small probabilities into addition, and it also allows us to calculate the derivatives more easily during the optimization process.  

\textit{Once fully trained, a generator component of the GAN is capable of converting random noise into new data samples (e.g., boots) that conform to the original distribution of the training data.} Essentially, the generator, without the use of a discriminator, can be inferred to obtain new data samples. GANs have been shown to be useful in a variety of applications such as image and text generation, data augmentation, and anomaly detection \cite{XIA2022497,10.1007/978-3-030-20893-6_39,deecke2018anomaly}. Naturally, this has spawned interest in the quantum information science community to develop corresponding quantum GANs.

\vspace{2mm}

\textit{Quantum GANs follow a similar generator and discriminator structure as classical GANs but use quantum principles to train and infer from the model.} Typically, the discriminator functions on classical resources, and the generator is trained on quantum resources. This is expected, as this structure allows the Quantum GANs to leverage quantum resources for generation tasks -- the component that persists beyond training; recall that the discriminator is essentially a quality inspector that can be decoupled after training and is not required during inference. While this structure is intuitive and has been demonstrated to be somewhat promising~\cite{huang2021experimental}, there are multiple open challenges that need to be overcome to achieve higher quality. 

\begin{figure}[t]
\centering
\includegraphics[scale=1.07]{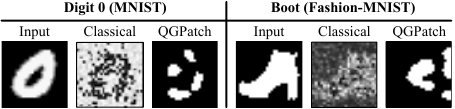}
\vspace{3mm}
\hrule
\vspace{1mm}
\caption{State-of-the-art quantum image generator, \sota{}~\cite{huang2021experimental}, generates low-quality images for different datasets. However, its quality is still similar to a classical GAN with $25\times$ more parameters and iterations.}
\label{fig:motivation}
\vspace{-5mm}

\end{figure}

\subsection{Limitations of Existing Quantum GANs}

A recent work by Huang et al.~\cite{huang2021experimental}, referred to as \sota{}, is the most complete, state-of-the-art demonstration of QuantumGANs on real quantum computers. \sota{} follows the Quantum GAN architecture described earlier but achieves limited effectiveness -- \sota{} can learn different shapes and produce recognizable images in some cases, but often suffers from low quality. Fig.~\ref{fig:motivation} shows that digit `0' is a reasonable approximation of the ground truth, but the generated boot image is far less recognizable. Despite the imperfect generations, we can see in Fig.~\ref{fig:motivation} how a classical GAN with $25\times$ the parameters of \sota{} and trained for $25\times$ as many iterations is worse or similar to the quantum technique -- supporting the potential of quantum machine-learning for image generation tasks as compared to the classical approach with much higher resources and complexity. This independently provides experimental evidence for why the quantum information science community is motivated to accelerate the progress of Quantum GANs, despite its current limitations.

\textit{The reasons for the limited effectiveness of Quatum GANs are multi-fold.} The quantum generator component runs on the quantum hardware and requires many qubits to produce high-quality images from random input noise (a qubit is the fundamental unit of computation and information storage on quantum computers). \sota{} addresses this challenge by breaking the image into ``patches'' and employing a generator for different patches. While this is reasonable for smaller resolution images, the ``patch-based'' approach suffers from scalability due to its fundamental nature of learning pixel-by-pixel. For example, a total of 245 qubits are required for images in the full-resolution 784-pixel MNIST dataset. 

\textit{The second challenge is performing efficient learning from random input noise -- it is critical for the quantum generator to effectively utilize the random input noise to generate a variety of images for the same class.} Inability to generate a variety of images within the same class -- which even classical GANs suffer from -- is popularly known as the ``mode collapse'' problem~\cite{problemsGAN}. Mode collapse is a side-effect of the generator learning to generate only produce one type of image for a given class because the generator has learned to ``fool'' the discriminator for this image and saturates in its learning. It is non-trivial to achieve high-quality generating while also maintaining variety. Motivated by these limitations, \sol{}'s design pushes the state of the art.

\subsection{Overview of \sol{} and Key Ideas}

\noindent\textbf{Hybrid Quantum-Classical Architecture.} \sol{} uses a hybrid quantum-classical architecture for image generation, as summarized in Fig.~\ref{fig:classical_and_quantum_gan}(b). The generator is quantum in nature and trained using quantum simulation, and the discriminator is classical -- similar to the construction in \cite{nakaji2021quantum, huang2021experimental}. However, there are two key novel architectural changes: (1) to address the scalability and quantum resource  bottlenecks for generating high-quality images, \sol{} applies a transformation on the input image dataset, and (2) \sol{} employs quantum-style random noise as input to enhance the ``variety or mode-collapse''  challenge. These two key features are described after a brief description of \sol{}'s generator and  discriminator.

\vspace{2mm}

\noindent\textbf{\sol{}'s Quantum Generator Network.} \sol{}'s quantum generator component is a network of multiple sub-generators. Each sub-generator is a \textit{variational quantum circuit} that is iteratively optimized to train a model. Variational quantum circuits are specific types of quantum circuits where some components of the circuits are tunable parameters, which are iteratively optimized. As a background, a quantum circuit is essentially a sequence of quantum gates applied to qubits. The quantum gates are fundamental operations that manipulate the qubit state, and any gate can be represented as a unitary matrix. One-qubit gates, such as the Pauli gates ($\sigma_x, \sigma_y, \sigma_z$), apply a rotation to just one qubit and are categorized by the axis around which the rotation takes place, and by how much. Multi-qubit gates, such as the CNOT gate, allow the creation of entanglement. 

\begin{figure}
    \centering
    \includegraphics[scale=0.46]{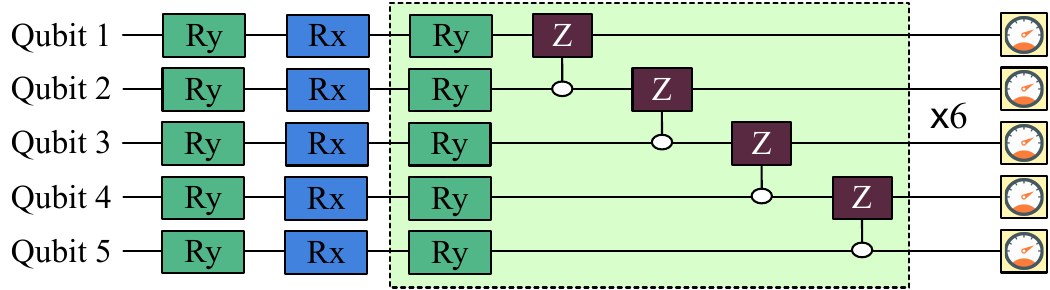}
    \vspace{1mm}
    \hrule
    \vspace{1mm}
    \caption{Circuit ansatz design of \sol{}'s generators. The RY gates in each layer are optimized during training.}
    \vspace{-5mm}
    \label{fig:circuit}
\end{figure}

Variational quantum circuits in \sol{} are comprised of two types of sections. The first type, $X$, has fixed gates that entangle the qubits. The second type, $\theta$, has tunable $U_3$ gates that are optimized via training. The overall circuit $V$ is therefore an optimization function built of unitary transformations such that $V = U(X, \theta)$. One key feature of variational quantum circuits is the ability to leverage many off-the-shelf classical deep-learning techniques for training and optimization.  This includes learning through common loss functions such as $L_2$ loss and leveraging existing optimizers such as Adam \cite{https://doi.org/10.48550/arxiv.1412.6980} that \sol{} leverages.

All sub-generator circuits in \sol{} have identical architectures, composed of a five-qubit circuit. Fig.~\ref{fig:circuit} shows an example of the sub-generator circuit. The circuit consists of encoding the input noise as angles using $Rx$ gates and $Ry$ gates. Following the embedding of the noise, the parameterized weights are encoded on each quantum layer alongside $CZ$ gates used to entangle the qubits at each layer. These weights contain the portion of the circuit that is optimized. Following these repeated layers, the $PauliX$ expected value is taken for each qubit in measurement. 

We note the simplistic design of \sol{}'s generator circuits is intentional; limiting the number of tunable parameters and depth of the circuits enables \sol{} to mitigate hardware error impact on real quantum machines and maintain high quality, as confirmed by our evaluation (Sec.~\ref{sec:eval}).

\vspace{2mm}

\noindent\textbf{Scalable Learning by Extracting Principal Components.} Unlike previous approaches, which learn to construct the pixels of an image directly and consequently, suffer from the scaling bottleneck, \sol{} demonstrates that extracting and learning the rich features is effective. The  principal component analysis (PCA) method enables efficient learning by focusing on the important features that compromise unique images, as opposed to having to learn to distinguish the important features from the redundant features in an image via the pixel-by-pixel method. PCA maximizes the entropy for a given number of components, concentrating the information of a dataset efficiently.

The first step in learning principal components is normalizing the input data. Once the images are normalized, \sol{} decomposes the images into principal components and scales these components between $[0,1]$ so that the quantum sub-generators can generate throughout the entire space of inputs. As quantum machines rely on only unitary transformations, it is not possible to obtain an output with an absolute value greater than one in measurement.  These scaled components are fed into the discriminator as the ``real labels'' and the generator begins to learn how to mimic the distribution throughout the training process. After learning, the outputs are scaled back to the original principal component range, then are inverse transformed to form a unique image that is within the distribution of the original images. This method helps us achieve higher quality in a scalable way. However, experimentally we found that this alone is not sufficient. \sol{} performs intelligent distribution of features to make the learning more effective and robust.

\vspace{2mm}

\begin{figure}
    \centering
    \includegraphics[scale=0.45]{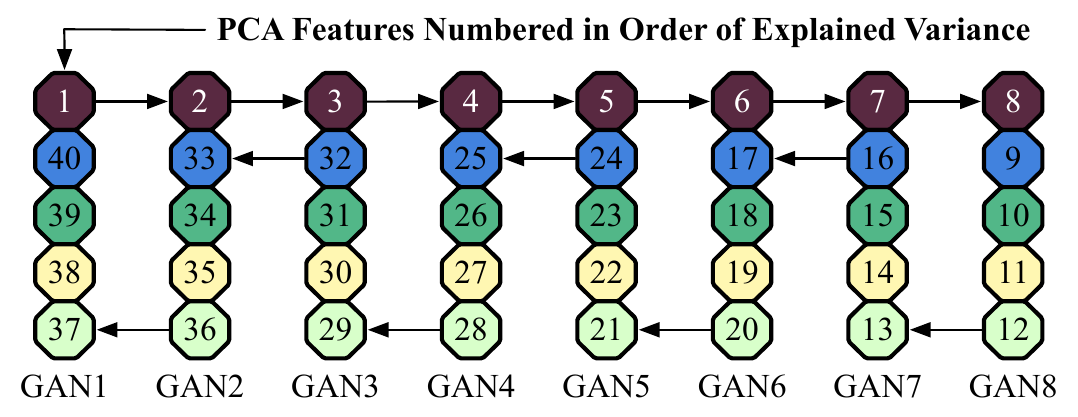}
    \vspace{0mm}
    \hrule
    \vspace{1mm}
    \caption{\sol{} divides the PCA features among the eight generator circuits in a manner that the total explained variance is close to equal across all generators.}
    \label{fig:reorder}
\end{figure}

\noindent\textbf{Feature Distribution Among Learners.} In the conventional implementation of the previously proposed idea, all generated features are distributed among several sub-generators  before being concatenated. Recall that the \sol{}'s quantum generator is an ensemble of multiple sub-generators, where each sub-generator is a variation quantum circuit. In the default case where PCA features are aligned one after another to the sub-generators, an unbalanced distribution of explained variance may emerge, where some sub-generators are responsible for significantly more important generations than others. This is because, by definition, the explained variance of PCA features is heavily concentrated on the first few features, and there is often a significant drop-off in explained variance from one PCA feature to the next. Recall that PCA features are entangled within sub-generators in training to form rich connections for image generation, when some sub-generators do not contain much useful information in any feature, the entire generator does not pose much utility. 

Therefore, while effective in achieving scalability, only performing PCA may not achieve the full potential of \sol{}'s quantum image generation as this leads to unbalanced training, likely to quickly discover gradients that ignore many sub-generators and plateaus in training. To mitigate this challenge, we propose a new PCA feature distribution mechanism to counteract the unbalanced nature of assigning principal components to sub-generators (Fig.~\ref{fig:reorder}). 

We begin by assigning the first principal component to the first sub-generator, the second principal component to the second, and so on until each sub-generator has one top principal component.  We follow this up by assigning the last n-1 principal components to the first generator, where n is the size of each sub-generator.  This is followed by assigning the second to last n-1 features to the second sub-generator and repeated on the third generator and so on until all features have been assigned.  This creates a much more balanced distribution, which enables us to ensure all sub-generators hold utility during the training process and the learning is not skewed. As an important side-note, \sol{}'s distribution also mitigates the pathological side-effects of hardware errors on NISQ quantum machines where the majority of critical principal component features could be concentrated on a single sub-generator. This sub-generator could be mapped to qubits with higher hardware error rates, and this can potentially make the training less effective. Our evaluation confirms the effectiveness of \sol{}'s principal component feature distribution among learners.  Finally, we describe how \sol{} utilizes the adaptive noise during image generation to increase the variety of images generated to avoid mode collapse.

\vspace{2mm}

\noindent\textbf{Adaptive Input Noise to Improve Variety of Generated Images.}  Recall that it is critical for the quantum generator to effectively utilize the random input noise to generate a variety of images for the same class. The inability to generate a variety of images within the same class leads to the ``mode collapse'' problem. Therefore, while \sol{}'s previous methods help us achieve high quality, it is critical to achieving high-quality, while also maintaining variety. 

\begin{figure}
    \centering
    \includegraphics[scale=0.46]{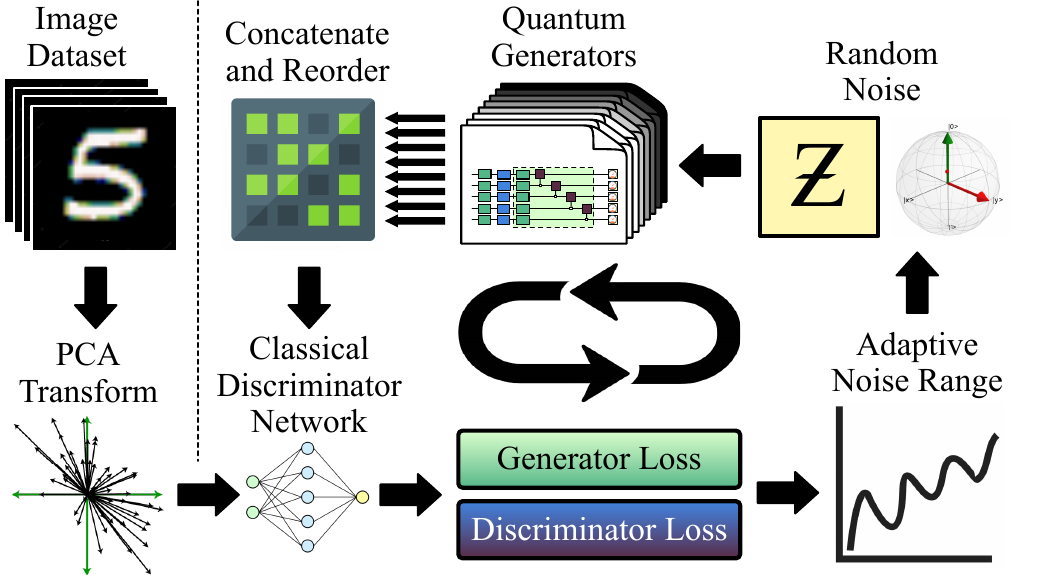}
    \vspace{3mm}
    \hrule
    \vspace{1mm}
    \caption{\sol{}'s process for training and optimizing the quantum generator circuits using PCA transformation, feature distribution, and adaptive noise generation.}
    \label{fig:mosaiq_training}
\end{figure}

To address this challenge, \sol{} introduces an adaptive noise range based on the ratio of training loss between the quantum generator and the classical discriminator. The noise range is determined by  the current progress of the generator discriminator mini-max game, instead of the traditional fixed range of $[0, \frac{\pi}{2}]$ as employed by \sota{}. Formally, the adaptive noise ($\text{Noise}_{\text{adaptive}}$) is defined below in terms of the Generator loss $G_L$ and Discriminator loss $D_L$ and the Ratio $\frac{G_{L_{0}}}{D_{L_{0}}}$ is the Discriminator/Generator loss ratio observed after the first epoch during training.
\begin{equation}
    \text{Noise}_{\text{adaptive}} = \dfrac{\pi}{8}  + \dfrac{5\pi}{8} \textit{ReLU}(\textit{Tanh}\big(\dfrac{D_L}{G_L} - \frac{G_{L_{0}}}{D_{L_{0}}}\big))
\end{equation}
As the above formula indicates, this adaptive bound enforces a minimum noise range of $\frac{\pi}{8}$ and a maximum range of $\frac{3\pi}{4}$ for the noise. While the bounds are chosen to be a specific constant, the noise is always distributed within this range and the upper bound automatically adjusts itself based on the relative effectiveness of the generator and the discriminator. \sol{} does not require tuning the thresholds for the range for different classes, it automatically adjusts itself to adapt to different conditions during training. 

\sol{} embeds the adaptive noise using angle embedding on the quantum circuit. The $Tanh$ function is used to scale the adaptive noise to asymptotically approach the range $[-1,1]$.  \sol{} chose $Tanh$ because it is a widely-used activation function in deep learning applications which transforms unbounded data into data bound by $[-1,1]$. $Tanh$ is defined as $Tanh(x) = \frac{e^{2x}-1}{e^{2x}+1}$. In practice, we observed that it is not effective to have a noise range lower than $\frac{\pi}{8}$, we leverage $ReLU$ to force negative values to be zero. $ReLU$ is also a commonly used activation function, however, $ReLU$ is used to provide a non-linear transform that makes all negative values 0 (to ensure that the noise is never less than $\frac{\pi}{8}$), and keeps all positive values the same.

While the non-linearity of $Tanh$ and $ReLU$ are important when used typically as activation functions, the non-linearity is not the main function in this application. If the generator is not able to catch up with the discriminator for some iterations at this small range, it should learn this smaller range instead of getting even smaller. We set the minimum as $\frac{\pi}{8}$ as we discovered that a lower value than $\frac{\pi}{8}$ results in mode collapse, with almost no distinction between images generated for a given class.

Leveraging adaptive noise for input ensures more consistent training as the generator can focus on quality when it is doing relatively worse compared to the classical discriminator and expand to generating more variety in high quality as it begins to fool the discriminator more on an easier objective. Overall, the adaptive noise mechanism helps us increase the variety of images for a given class while ensuring that the image quality also remains high. This is further confirmed by our evaluation (Sec.~\ref{sec:eval}).

\vspace{2mm} 

\noindent\textbf{\sol{}'s Classical Discriminator.} \sol{} 's discriminator is a classical deep-learning network, designed to aid in the training of the quantum generator. This network is essentially training wheels to guide a difficult-to-train quantum GAN, which can be discarded after training. The discriminator is much larger than the quantum generator, as there is only a single discriminator that must compete in the adversarial game with multiple quantum generators. Typical to its classical counterpart, \sol{}'s discriminator has multiple linear layers with $ReLU$ activation (e.g., 64 layers). \sol{} has a terminal layer with $Sigmoid$ activation which introduces more non-linearity into the network, which enables it to learn more complex structures. The single value produced at the end of the discriminator network allows it to act as a classifier to distinguish between real and generated data.

\vspace{2mm}

\begin{figure}
    \centering
    \includegraphics[scale=0.51]{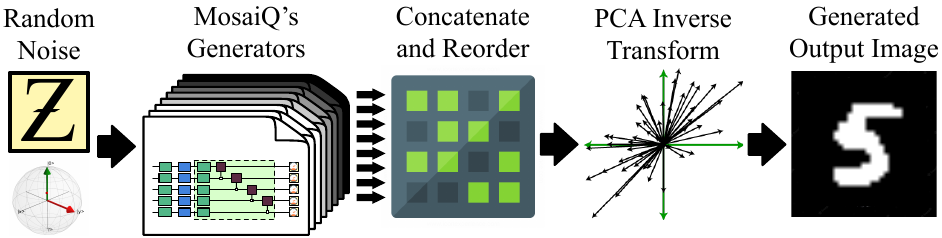}
    \vspace{3mm}
    \hrule
    \vspace{1mm}
    \caption{\sol{}'s inference process to generate new images via running the generators on quantum machines.}
    \label{fig:mosaiq_inference}
\end{figure}

\noindent\textbf{Putting It All Together.} The overall workflow for training and inference for \sol{} are visually summarized in Fig.~\ref{fig:mosaiq_training} and~\ref{fig:mosaiq_inference}, respectively. Fig.~\ref{fig:mosaiq_training} depicts the continuous feedback-driven process where the discriminator and generator participate in a non-cooperative game to improve the overall quality of \sol{}. Depending upon the relative loss of the discriminator and generator networks, the random input noise is automatically adjusted to help \sol{} generators avoid the problem of mode collapse problem and enable it to generate variety. During the training process, the input images are transformed using PCA to encode critical and more information in a compact manner for resource-limited NISQ quantum computers, instead of learning over the input images pixel by pixel. Finally, the five-step inference procedure shown in Fig.~\ref{fig:mosaiq_inference} highlights that the PCA transformation is \textit{inverted} to generate new images when running the fully-trained \sol{} generators on real quantum computers for inference to generate new images.

\section{Experimental Methodology}
\label{sec:eval}
\noindent\textbf{Datasets.} \sol{} is evaluated on the MNIST \cite{mnist} and Fashion MNIST datasets~\cite{xiao2017/online} as they have been widely-used for QML evaluation on NISQ-era quantum machines~\cite{nakaji2021quantum, huang2021experimental}. MNIST consists of 28x28 gray-scale images of handwritten digits from 0 to 9. Fashion MNIST consists of 28x28 gray-scale images of clothing and accessories. Fashion MNIST provides more challenges for image generation and is chosen to explore more domains of quantum image generation. For both datasets, we split the dataset by image label and train individual models for each data type. This technique has been used in \cite{huang2021experimental} and allows us to interpret the generation process and difficulties isolated for each class of data.

\vspace{2mm}

\noindent\textbf{Experimental Framework and Training Details.} The environment for \sol{} consists of PyTorch \cite{NEURIPS2019_9015} acting as a wrapper for Pennylane \cite{https://doi.org/10.48550/arxiv.1811.04968}. \sol{} is trained on IBM's quantum simulator for speed, but the inference is performed on both the quantum simulator and real quantum machine. For real quantum machine experiments, Pennylane compiles the circuits into a backend compatible with IBM-Qiskit \cite{Qiskit}. All real machine runs were performed on the IBM QX Jakarta machine \cite{Qiskit}. Images used in training are selected from the training set and decomposed to principal components of size 40. These components are divided across eight five-qubit sub-generators. The discriminator learns to differentiate at the level of principal components and does not need to utilize a full image. Our final FID metrics reported are at the end of the training after 500 iterations (where all methods achieve near-final stability). 

\sol{} uses PyTorch's Standard Gradient Descent Optimizer for both the generator and discriminator and Binary Cross Entropy Loss for the shared loss of the generator and discriminator. The Adaptive Noise range is simple to calculate based on the generator and discriminator loss and \sol{} automatically guides it to adjust itself every training iteration. The generator learning rate is $0.3$ and the discriminator learning rate to $.05$ with a batch size of $8$. 

\vspace{2mm}

\noindent\textbf{Framework for Competing Techniques.} We set up \sota{}, the state-of-the-art technique~\cite{huang2021experimental}, using the popular Pennylane framework and choose the parameters used in the original paper~\cite{huang2021experimental}.
Our \sota{} training is performed using the batch size of eight and it trains until the quality stabilizes (500 iterations). We use the same network size that is mentioned in the original paper, using 4 sub-generators with 5 features each.  Evaluating the probabilities in measurement yields a 64-pixel (8x8) image. As \sol{}'s results are based on the original size of the datasets of 784 pixels, we upscale the results of \sota{} using Bilinear Interpolation to allow for direct comparison. We rigorously explored multiple interpolation techniques to provide \sota{} as much as an advantage as possible and experimentally determined that Bilinear interpolation yielded the highest quality upscaling. For \sota{}, similar to \sol{}, all metrics scores shown are calculated based on 500 images generated at the end of training compared to the entire distribution of the respective image category.

We additionally set up other classical experiments to act as ablations for the efficacy of \sol{}. We introduce two techniques: (1)  PCAInverse (using random inputs and PCA inverse transformation to generate images), and (2) ClassicalPCA (a purely-classical GAN using the same number of parameters as MosaiQ, but uses MosaiQ’s PCA technique for feature compression). PCAInverse applies the same inverse PCA transformation as \sol{}, to random noise of size 40. This technique is designed to explore the isolated effects of the inverse PCA transformation procedure since it does not perform any learning. ClassicalPCA trains a purely-classical GAN with an identical number of parameters as \sol{} and apply the same inverse PCA transformation workflow.

\vspace{2mm}

\begin{figure}[t]
\centering
\includegraphics[scale=0.51]{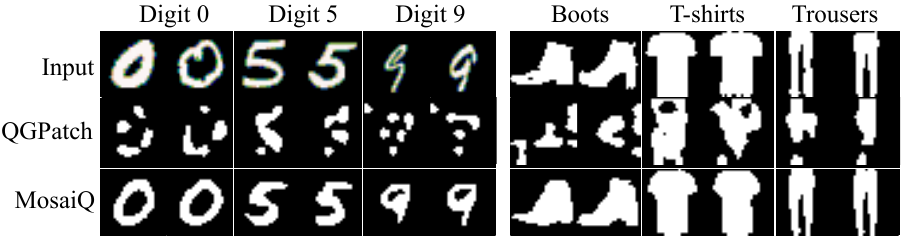}
\vspace{1mm}
\hrule
\vspace{1mm}
\caption{\sol{} produces visually higher-quality images compared to the state-of-the-art technique, \sota{}, for different classes of MNIST and Fashion-MNIST datasets.}
\label{fig:image_vis_mnist}
\vspace{-8mm}
\end{figure}

\noindent\textbf{Figures of Merit.} To capture the quality of image generation, our evaluation figures of merit are both qualitative and quantitative.  Our primary quantitative figure of merit is FID (Fréchet inception distance) score \cite{heusel2017gans}. FID evaluates the distance between two distributions, as defined below for a Gaussian with mean $(m,C)$ and a second Gaussian with mean $(m_w, C_w)$: $\text{FID}=  \norm{m-m_w}_2^2 + Tr(C + C_w - 2(CC_w)^{1/2})$. A lower FID score between two distributions indicates higher similarity -- and hence, lower FID scores imply higher quality.  The FID score has been shown to provide a more meaningful comparison over alternative metrics (e.g., Inception Score) for image GANs~\cite{heusel2017gans}. 

We also evaluate the variance of our images, by evaluating the variance of the pixel values relative to the mean.  As defined below, the variance metric gives insight into the distinctness of the images generated for each method.  For each pixel in an image row $r$ and column $c$ in an image, we sum the squared difference from the mean value for that respective pixel to achieve our variance score $V$. We then take the cumulative density function (CDF) of these variance scores for each image generated image $G(z)$ given uniformly distributed noise $z$ from $[0,\pi/2]$: $V = \sum_{r}\sum_{c} (\mu_{rc}-G(z)_{rc})^2$. A higher variance indicates higher variety in generated images for a given class -- which is desirable, however, it is also critical that a method should attain a lower FID score before demonstrating higher variety. 
\section{Evaluation and Analysis}

In this section, we go over the results of \sol{} and provide an analysis of the key elements of the design.

\vspace{2mm}

\begin{figure}[t]
\centering
\includegraphics[scale=0.52]{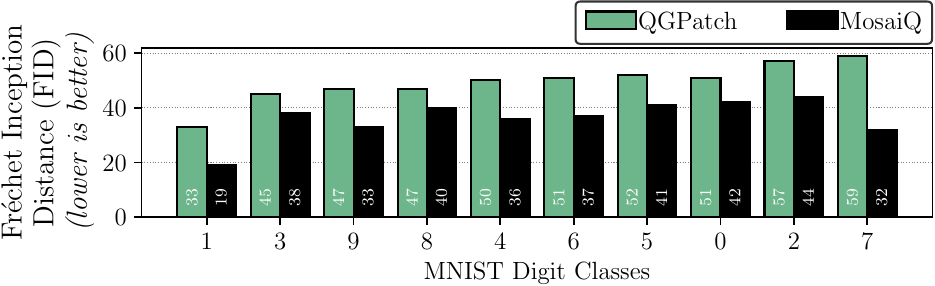}
\vspace{1mm}
\hrule
\vspace{1mm}
\caption{\sol{} consistently produces lower FID score images (\textit{higher quality images}) compared to \sota{} across different classes of the MNIST dataset.}
\label{fig:image_fid_mnist}
\end{figure} 

\begin{figure}[t]
\centering
\includegraphics[scale=0.52]{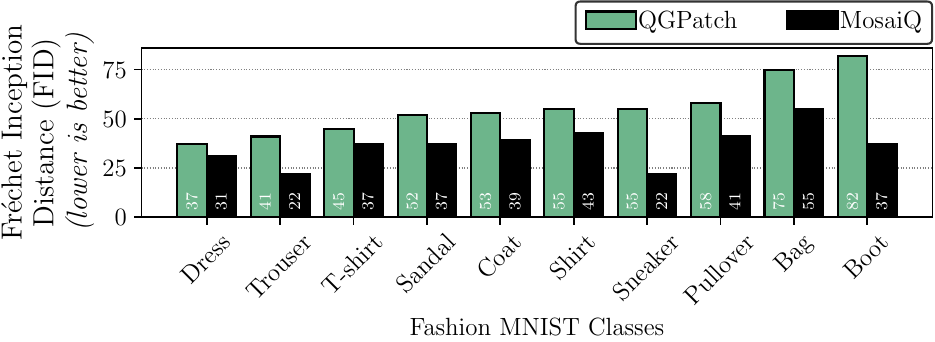}
\vspace{1mm}
\hrule
\vspace{1mm}
\caption{\sol{} consistently produces lower FID score images (\textit{higher quality images}) compared to \sota{} across different classes of the Fashion-MNIST dataset.}
\label{fig:image_fid_fashion_mnist}
\end{figure} 

\begin{figure}[t]
\centering
\includegraphics[scale=0.51]{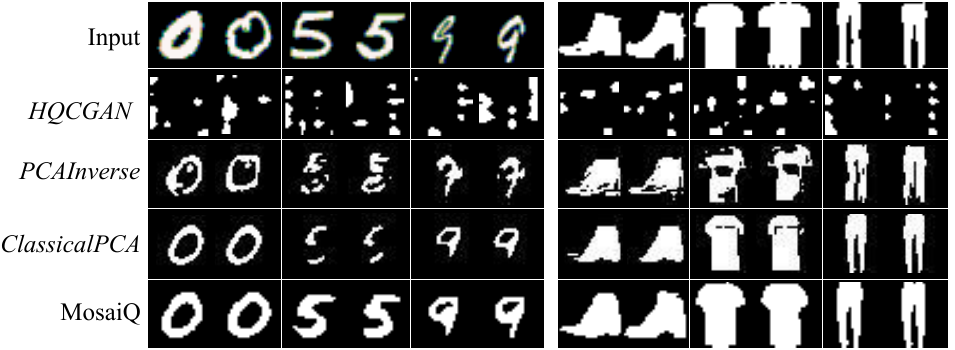}
\vspace{1mm}
\hrule
\vspace{1mm}
\caption{\sol{} produces higher quality images on the MNIST and Fashion MNIST dataset as compared to HQCGAN \cite{tsang2022hybrid}. \sol{} produces higher quality images than the classical methods: PCAInverse and ClassicalPCA.}
\label{fig:hqcgan}
\end{figure}

\begin{figure}[t]
\centering
\includegraphics[scale=0.52]{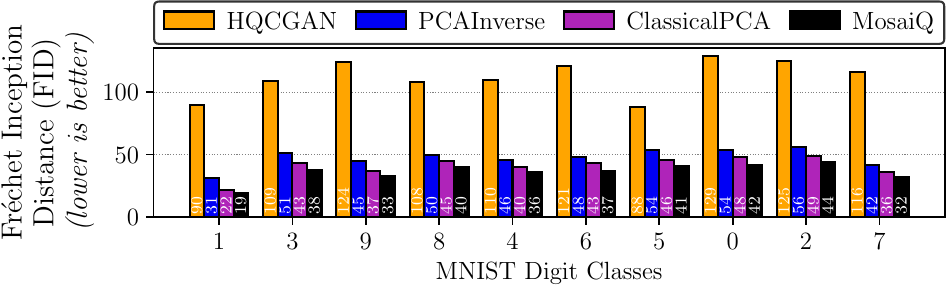}
\vspace{1mm}
\hrule
\vspace{1mm}
\caption{\sol{} produces higher quality images on every class of the MNIST dataset as compared to HQCGAN \cite{tsang2022hybrid} in addition to the classical methods tested.}
\vspace{-2mm}
\label{fig:fidhqcgan}
\end{figure}

\begin{figure}[t]
\centering
\includegraphics[scale=0.52]{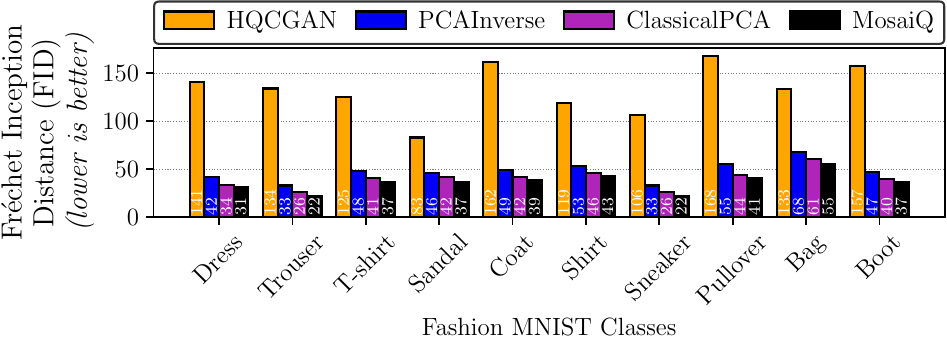}
\vspace{1mm}
\hrule
\vspace{1mm}
\caption{\sol{} produces higher quality images on every class of the Fashion MNIST dataset as compared to HQCGAN \cite{tsang2022hybrid} in addition to the classical methods tested.}
\vspace{-2mm}
\label{fig:fidhqcganfashion}
\end{figure}

\noindent\textbf{\sol{}: Key Results and SOTA Comparison.} \sol{} yields significantly better image quality both visually and quantitatively compared to the state-of-the-art (SOTA) method, \sota~\cite{huang2021experimental}. As discussed earlier, \sol{} is evaluated on two different datasets to cover diversity across datasets and within a dataset with different shapes and styles (digits and clothing). First, we highlight that \sol{} provides a significant visual enhancement in the images produces across different classes compared to the SOTA method. In fact, Fig.~\ref{fig:image_vis_mnist} shows that the SOTA-produced images are almost unrecognizable for sophisticated shapes (`0' vs. `5' and different types of shoes). In comparison, \sol{} is effective across different class types -- this is further substantiated by lower FID scores of the images generated by \sol{} (Fig.~\ref{fig:image_fid_mnist} and~\ref{fig:image_fid_fashion_mnist}). 

As expected, the FID score obtained by \sol{} varies across different classes for both datasets; this is because of varying degrees of difficulty and shapes of different images. However, the most notable observation is that \sol{} consistently outperforms \sota{} across all class types by a significant margin (by more than 10 points in many cases). In fact, MNIST digit `7' is quite challenging for \sota{} (FID score is 60), and \sol{} improves the FID score by approximately 20 points. Similarly, the most challenging class in the Fashion-MNIST dataset (Boot) receives a 45-point improvement by \sol{}. The key reason \sol{} outperforms \sota{} is the efficiency of principal components in capturing information.  Instead of having to learn pixels one by one, \sol{} scales more easily by learning a distribution of features that better organize redundancy.  For example, most of the background of MNIST and Fashion MNIST images are black, in the case of \sota{}, each pixel must be synthesized, where \sol{} may be able to build these features with only a few principal components.

We also compare \sol{} to other methods, including the PCAInverse technique, the ClassicalPCA technique, and a recent work on hybrid quantum GANs which we refer to as HQCGAN \cite{tsang2022hybrid}. We sample generated images in Fig. \ref{fig:hqcgan} and find the images produced by all methods tested are lower quality than \sol{}, producing images which are far less human-recognizable. We show the corresponding FID scores for MNIST in Fig \ref{fig:fidhqcgan} and Fashion MNIST in Fig. \ref{fig:fidhqcganfashion}.  We find that \sol{} produces higher quality images and significantly improved FID scores in all cases tested compared to the classical methods (PCAInverse and ClassicalPCA), and HQCGAN. Importantly, \sol{} outperforms ClassicalPCA, demonstrating the power of quantum networks in image generation when compared to equal-sized classical networks. Our results show that while HQCGAN is promising and useful, the final quality may not be as high as MosaiQ -- this is because HQCGAN requires significantly more parameters, more training resources, lacks noise-mitigation, and learns noise in a fixed range ($[0,1)$ instead of MosaiQ’s adaptive noise range technique). Next, we dissect the key reasons behind \sol{}'s effectiveness in more detail.  

\begin{figure}[t]
\centering
\includegraphics[scale=0.54]{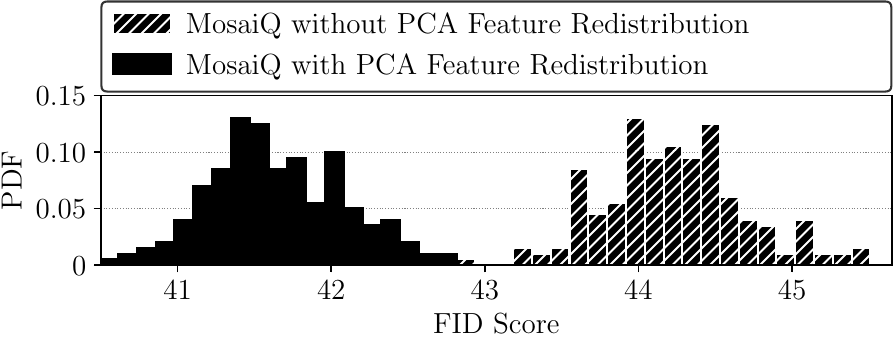}
\vspace{1mm}
\hrule
\vspace{1mm}
\caption{\sol{}'s PCA feature distribution among learners improves the quality of generated images. The figure shows a distribution of 200 FID scores (comparing 8 images each to the entire data distribution) for the case with and without PCA feature redistribution.}
\vspace{-4mm}
\label{fig:fid_pca_distribution}
\end{figure}

\vspace{2mm}

\noindent\textbf{Why Does \sol{} Work Effectively?} \textit{Effect of careful PCA feature distribution among sub-generators.} Recall that \sol{} employs an intelligent PCA feature distribution among sub-generators in the generation phase. The goal is to equalize the explainable PCA feature variance across learners -- in an effort to make learners equally capable instead of weaker learners not effectively contributing toward the overall quality. To better understand and demonstrate the effectiveness of this mechanism, Fig.~\ref{fig:fid_pca_distribution} shows the FID score over multiple training iterations with and without this mechanism while keeping all other design elements intact. For easier interpretation, Fig.~\ref{fig:fid_pca_distribution} shows this for class digit `5' for the MNIST dataset. We observe that PCA feature distribution allows \sol{} to achieve lower (and hence, better) FID score over training compared to without employing this mechanism. A side benefit of this mechanism is that \sol{} can also mitigate hardware errors on real quantum computers, as discussed later.

\begin{figure}[t]
\centering
\includegraphics[scale=0.54]{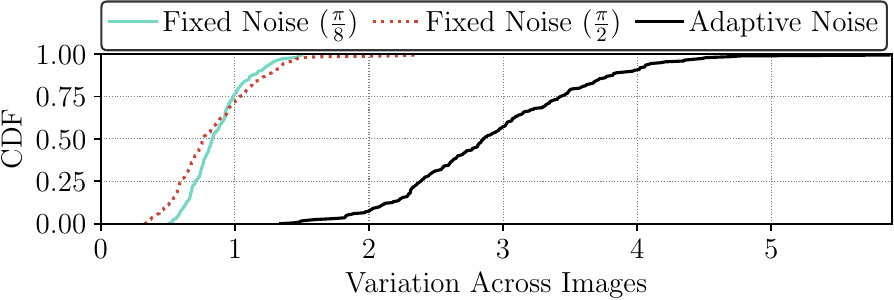}
\vspace{3mm}
\hrule
\vspace{1mm}
\caption{Adaptive noise during generation helps \sol{} achieve variety (higher variance across images for the same class to mitigate mode-collapse challenge), while \textit{achieving lower FID score at the same time, as shown earlier}.}
\label{fig:adpative_noise_variance}
\end{figure}

\vspace{2mm}

\noindent{\textit{Effect of adaptive noise generation during training.}} Fig.~\ref{fig:adpative_noise_variance} shows the effect of adaptive noise generation that is used by the \sol{} generators instead of using a constant noise threshold range (i.e., $ [0, \frac{\pi}{2}]$) used by the \sota{} method. Recall that GANs often suffer from intra-class mode-collapse challenges where they can produce a high-quality image for a given class, but there is a possibility that all generated images for a given class appear similar. Therefore, it is critical to ensure that the generated images for a given class have sufficient variety. Fig.~\ref{fig:adpative_noise_variance} confirms that adaptive noise improves variety (class digit `5' used as an example) compared to fixed noise ranges. This is because adaptive noise enables the generators to learn over different distributions more effectively and generate variations. Having an adaptive range allows the generator to improve variety when it is doing comparatively well and focus on stability when it is performing poorly with a smaller range of inputs.  Taking advantage of this allows us to have high stability in training and high variety over time. 
\begin{figure}[t]
\centering
\includegraphics[scale=0.59]{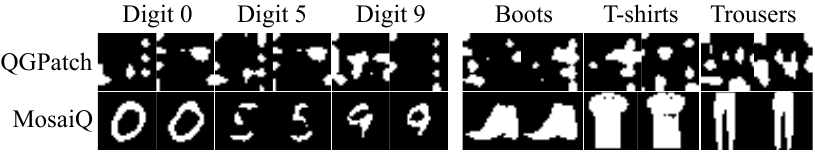}
\vspace{4mm}
\hrule
\vspace{0mm}
\caption{Visual image quality of images produced by \sol{} on real IBM quantum machine (Jakarta) for six representative classes for MNIST and Fashion-MNIST datasets. }
\label{fig:real_image_vis}
\vspace{-4mm}
\end{figure}

\vspace{2mm}

\noindent\textbf{\sol{} on NISQ Quantum Machines.} Our experimental campaign on real superconducting-based IBM NISQ quantum computers confirms \sol{} produces higher quality images across different datasets with diverse classes of images. Fig.~\ref{fig:real_image_vis} and Table~\ref{tab:real_fid_table} summarize the image quality results for \sota{} and \sol{}. While \sol{}'s quality remains consistent with simulations, \sota{}'s quality degrades considerably on real computers.  In fact, our results revealed \sota{} produces lower quality images and sometimes unrecognizable images because \sota{} is more prone to side-effects of prevalent quantum gate errors on real machines, while \sol{}'s simplistic design successfully mitigates those effects as discussed next. 

\begin{table}[t]
\begin{center}
\caption{\sota{} and \sol{}'s FID scores on a real quantum computer for MNIST \& Fashion-MNIST datasets.}
\label{tab:real_fid_table}
\vspace{-2mm}
\subfloat[\sota{}]{
\resizebox{\columnwidth}{!}{
\begin{tabular}{|c|ccc|ccc|} 
 \hline
 Environment & \ Digit 0 & \ Digit 5 & \ Digit 9 &\ T Shirt & \ Pants & \ Shoes \\[0.5ex] 
 \hline\hline
 Quantum Simulator & 52 & 52 & 48 & 45 & 42 & 82  \\ 
  \hline
 IBM Jakarta Machine & 145 & 134 & 131 & 125 & 144 & 124  \\ 
 \hline
\end{tabular}}}
\vspace{-2mm}
\subfloat[\sol{}]{
\resizebox{\columnwidth}{!}{
\begin{tabular}{|c|ccc|ccc|} 
 \hline
 Environment & \ Digit 0 & \ Digit 5 & \ Digit 9 &\ T Shirt & \ Pants & \ Shoes \\[0.5ex] 
 \hline\hline
 Quantum Simulator & 42 & 42 & 33 & 34 & 22 & 38  \\ 
  \hline
 IBM Jakarta Machine & 45 & 44 & 35 & 38 & 23 & 38  \\ 
 \hline
\end{tabular}}}
\end{center}
\vspace{-5mm}
\end{table}

First, we observe that \sol{} produces competitive quality images despite hardware errors on real quantum computers. Due to high queue wait time and limited availability of quantum resources, we are presenting results only for selected classes (three from each dataset) which cover a diverse range of intricacies in terms of shape and information. In fact, via visual inspection for different images produced for a given class, we observe that the variety in the produced images by \sol{} is also high on the quantum computer - effectively mitigating the mode-collapse pitfall in GANs. This is because of \sol{}'s adaptive noise mechanism during the generation phase that improves image variety and avoids mode collapse. 

Second, we observe that the FID scores of \sol{} on a real quantum computer are similar to error-free simulation results, across different classes. This result demonstrates that \sol{}'s design is effective even when hardware errors are present. In fact, the IBM Jakarta computer has a considerably low quantum volume \textit{(higher is better)} with a quantum volume of 16. Quantum volume is a useful metric in determining the error rate and capabilities of a quantum computer. For perspective, at the time of writing, IBM has computers available with quantum volumes of 128. We chose a relatively less capable quantum computer for our experiments to demonstrate the portability and robustness of \sol{}. The reason for \sol{}'s effectiveness is twofold: (1) its simplistic design for generator learners, which avoids a large number of two-qubit gates, and circuits with high depths. which are more prone to hardware noise, and (2) its PCA feature distribution mapping among the ensemble of learners in the generator -- this careful redistribution ensures that most critical PCA features are not concentrated on a few qubits, which in the pathological case could be most severely impacted by hardware-errors. \sol{}'s PCA redistribution ensures that the learners do not need to know the error rate of individual qubits on the computer or make assumptions about different error types.

\section{Related Work}

\noindent\textbf{Demonstration of the Speedup Achieved by Quantum GANs.} Quantum GANs were first introduced by Lloyd et al.\cite{PhysRevLett.121.040502}. This work theoretically proved the potential for exponential superiority of Quantum GANs over their classical counterparts. This was later followed by the work of Dallaire-Demers et al. \cite{PhysRevA.98.012324}, which established a way to design and train variational quantum circuits as the generator component of a Quantum GAN model. Nakaji et al.~\cite{nakaji2021quantum} use a hybrid configuration with a classical discriminator and a quantum generator for classification. Their results demonstrate the potential for quantum GANs to solve problems that are intractable with classical GANs.

\vspace{2mm}

\noindent\textbf{Enchancing Classical GANs using Quantum Components.} Approaches such as the work done by Rudolph et al.~\cite{rudolph2022generation} use a quantum enhancer for large-scale fully-classical generators to enhance the quality of image generation. While the images generated are high quality, this is mostly attributed to the large classical generators. The visual results in these works seem appealing because the classical GAN-based baseline is already high quality, with a quantum preamble which only improves the image quality slightly. This is in contrast to \sol{} which uses a quantum generator to generate images without a classical generator.

\vspace{2mm}

\noindent\textbf{Fully-Quantum Generators.} The work of Huang et al.~\cite{huang2021quantum} uses quantum gradients to train quantum GANs to replicate logic gates such as XOR with high fidelity and also generate images from the MNIST dataset. The technique relies on using quantum-based loss functions for their generator and discriminator. This work also shows that quantum GANs have similar performance as classical GANs while using 94.98\% fewer parameters. A succeeding work with higher quality MNIST image generation can be found in \sota{}~\cite{huang2021experimental}, with a generator split among several sub-generators each responsible for generating a part of the full image. \sota{} trains a quantum generator against a classical discriminator. The quality of generation is the highest among pre-\sol{} quantum GANs with a purely quantum generative component, at the time of writing. Therefore, we compare \sol{} against \sota{} in this paper and show that \sol{} outperforms \sota{} in terms of image quality and variety.

\section{Conclusion}

\sol{} is the first quantum GAN to successfully generate high-quality images on real quantum computers. MosaiQ incorporates PCA feature redistribution and adaptive noise to facilitate improved interaction between the quantum generator and classical elements of the model. This integration enhances the quality and variety of the generator's output respectively. 

\noindent\textbf{Acknowledgements} We thank the reviewers for constructive feedback. This work was supported by NSF Award 2144540, Northeastern University, and Rice University.

{\small
\bibliographystyle{ieee_fullname}
\bibliography{egbib}
}

\end{document}